\documentclass[conference]{IEEEtran}
\IEEEoverridecommandlockouts
\usepackage{cite}
\usepackage{amsmath,graphicx,amssymb,algorithm,algorithmic,url}
\usepackage{algorithmic}
\usepackage{graphicx}
\usepackage{textcomp}
\usepackage{xcolor}
\def\BibTeX{{\rm B\kern-.05em{\sc i\kern-.025em b}\kern-.08em
    T\kern-.1667em\lower.7ex\hbox{E}\kern-.125emX}}
\begin{document}

\title{A New Approach of Data Pre-processing for Data Compression in Smart Grids}

\author{Yifei Sun$^{\star}$ \qquad Hang Zou$^{\star}$ \qquad Samson Lasaulce$^{\star}$ \qquad Michel Kieffer$^{\star}$ \qquad Lucas Saludjian$^{\dagger}$\\
$^{\star}$ L2S, CNRS-CentraleSupelec-Univ. Paris Sud, Gif-sur-Yvette, France\\
	$^{\dagger}$ RTE, France}


\maketitle

\begin{abstract}
The conventional approach to pre-process data for compression is to apply transforms such as the Fourier, the Karhunen-Lo\`{e}ve, or wavelet transforms. One drawback from adopting such an approach is that it is independent of the use of the compressed data, which may induce significant optimality losses when  measured in terms of final utility (instead of being measured in terms of distortion). We therefore revisit this paradigm by tayloring the data pre-processing operation to the utility function of the decision-making entity using the compressed (and therefore noisy) data. More specifically, the utility function consists of an Lp-norm, which is very relevant in the area of smart grids. Both a linear and a non-linear use-oriented transforms are designed and compared with conventional data pre-processing techniques, showing that the impact of compression noise can be significantly reduced.
\end{abstract}

\begin{IEEEkeywords}
Compression, Learning, Neural Networks, Preprocessing, Transform.
\end{IEEEkeywords}

\section{Introduction}
\label{sec:Introduction}

Smart grids are, in particular, electricity networks equipped with many sensors to monitor the states of the grid (e.g., the voltage, the current, the phase, the temperature), which generates very large volumes of data to be compressed and transmitted through the grid. This strongly motivates the research in the area of compression of smart grid measurements (see e.g., \cite{6684605}). Signals measured in smart grids are quite specific and make classical compression techniques unsuited. This is why researchers had to adapt some existing schemes to the context of smart grids. Despite of that, the used mathematical tools to design good compression schemes are quite classical. For instance, in \cite{637001}-\cite{5664816} the discrete wavelet transform (DWT) is that the core of the proposed design. In \cite{7202904}, \cite{6469311}-\cite{6145362}, the singular value decomposition and principal component analysis (PCA) are respectively exploited. More advanced and more specific designs have been proposed e.g., in \cite{7286366} where the authors propose a real-time data compression scheme based on an algorithm combining exception compression with swing door trending compression. In \cite{7741464}, the authors propose to exploit some priors on the consumers' behavior to improve data compression. 

When inspecting all the quoted works and the underlying literature, it can be seen that the main objective is either to decrease the required data rate to reach a given compression quality (that is minimizing the distorsion) or to improve as much as possible the compression quality for a fixed rate. Recently, the authors of \cite{zhang-wiopt-2017}\cite{8629632} have questioned this point of view. Indeed, if one assumes that the decision-making entity has to make a binary decision, having at its disposal high quality data measurements might be useless and, worse that that, it may involve costly system overdimensioning (in terms of storage space, bandwidth, processing capability, etc). Rather, the new paradigm which is suggested in \cite{zhang-wiopt-2017}\cite{8629632} is to consider the final use made by the decision-maker using the compressed data. When this use can be modeled as the maximization of a given utility function, it becomes possible to taylor the compressor to the final use. In \cite{zhang-wiopt-2017}\cite{8629632}, the authors focus on the quantization stage of the compressor and the data use consists in choosing a power control vector based on a compressed version of the channel state. In the present paper, we develop for the first time this new paradigm for the problem of data processing and consider as the final use the Lp minimization problem, which is fundamental in smart grids since it encompasses key problems such as the peak power, total consumed energy, and Joule losses minimization problems.

The paper is structured as follows. In Section \ref{sec:Problem-formulation} we formulate the problem to be solved. In particular, we define the final utility function to be maximized and introduce the proposed optimality loss measure. In Section \ref{sec:Solution-for-a-fixed-l} we provide an approximation of the solution of the problem of Section \ref{sec:Problem-formulation}. This approximation is very useful to derive the best linear use-oriented pre-processing technique in Section \ref{sec:linear_transformation}. In \ref{sec:nonlinear_transformation}, we exploit a neural network to design a non-linear use-oriented pre-processing technique and there to be able to assess the benefits from having a non-linear transform instead of a linear one. Numerical results are provided in Section \ref{sec:Simulation} and conclusions are given in Section \ref{sec:conclusion}.

\section{Problem formulation}
\label{sec:Problem-formulation}
In this paper, we assume that the use of the decision-making entity (e.g., a home flexible consumption power scheduler, an electric vehicle battery charging controller, a hot-spot temperature distribution transformer controller, a transmission network flow controller) using the compressed data can be modeled as the maximization operation of a certain utility function. The considered utility function expresses as follows: 
\begin{equation}
u\left(x;\ell\right)=-||x+\ell||_{p}\label{eq:Utility}
\end{equation}
\begin{itemize}
	\item where the Lp-norm of a generic vector $v$ of size $n$ is defined by $\|v\|_p=(|v_1|^p+\dots+|v_n|^p)^{1/p}$;
	\item $x=\left[x_{1},x_{2},\dots,x_{P}\right]^{T}\in\mathbb{R}_{+}^{P}$ corresponds to the decision to be taken by the decision-maker;
	\item  $\ell=\left[\ell_{1},\ell_{2},\dots,\ell_{P}\right]^{T}$ corresponds to the parameters of the function to be maximized and precisely correspond to the data to be measured and compressed. This means that the decision-maker has in fact only access to a compressed version of $\ell$, which will be denoted by $\widehat{\ell}$. 
\end{itemize}
In the case of a home consumption power scheduler, $x$ would represent the flexible consumption vector to be chosen and $\ell$ would represent the non-controllable or non-flexible part of the consumption, and only a compressed version of it would be available to the scheduler. With $p=1$, the $x$ maximizing $u$ minimizes the total energy consumption. For $p=2$, it minimizes Joule losses (whenever $x$ and $\ell$ are interpreted a currents). For $p\rightarrow \infty$, one minimizes the peak power. 

If one assumes that $x_i$ is a power and one has to fulfill a given energy need $E$, the optimization problem (OP) of interest associated with the task to be executed by the decision-maker becomes: 
\begin{equation}
\begin{aligned}   \text{maximize} &\ u\left(x;\ell\right)\\
\text{s.t.}\  & \sum_{j=1}^{P}x_{j}-E=0\\
& x_{j}\geqslant0,\ j=1,\dots,P,
\end{aligned}
\label{eq:xstar}
\end{equation}
for any $E>0$. $x^\star\left(\ell\right)$ is one solution for this OP, and we consider $\ell\in\mathbb{R}^{P}$ for the reason as follows.

In order to reduce the volume of the data to transmit, a lossy compression technique is implemented. Thus the exact value of $\ell$ are usually unavailable to make a decision.

For this reason, when solving \eqref{eq:xstar}, assume that only an approximate value
$\widehat{\ell}\in\mathbb{R}^{P}$ of $\ell$ is available.
Replacing $\ell$ by $\widehat{\ell}$ in
\eqref{eq:xstar}, one obtains $x^{\star}\left(\widehat{\ell}\right)$.

The resulting utility becomes
\begin{equation}
u\left(x^{\star}\left(\widehat{\ell}\right);\ell\right)=-||x^{\star}\left(\widehat{\ell}\right)+\ell||_{p}.
\end{equation}

Due to the approximate value $\widehat{\ell}$ obtained by lossy compression, $u\left(x^{\star}\left(\widehat{\ell}\right);\ell\right)\leqslant u\left(x^{\star}\left(\ell\right);\ell\right)$. And an utility loss is caused.

Our objective is to design a source coding scheme which minimizes
the utility loss.
\begin{figure}[h]
	\includegraphics[scale=0.39]{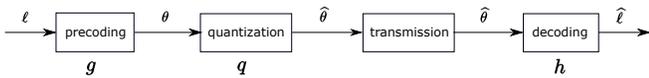}
	
	\caption{Coding scheme}
	
	\label{fig:Coding-scheme}
\end{figure}

Fig. \ref{fig:Coding-scheme} illustrates this coding scheme.

Precoding scheme can be considered as a function
\[
\begin{array}{cccc}
g: & \mathbb{R}_{+}^{P} & \rightarrow & \mathbb{R}^{N}\\
& \ell & \mapsto & \theta
\end{array}
\]
where $N\leqslant P$. By precoding function $g\left(\cdot\right)$, the encoded parameters, $\theta\in\mathbb{R}^{N}$, are obtained.

Because of the bit constraint, quantization precessing follows after the precoding. Assume that a quantization function $q\left(\cdot\right)$ is available,
thus one obtains quantized encoded parameters $\widehat{\theta}=q\left(\theta\right)$. And $\widehat{\theta}$ is transmitted losslessly.

The reconstructed signal is obtained by the quantized encoded parameters and a decoding function $h\left(\cdot\right)$,  $\widehat{\ell}=h\left(\widehat{\theta}\right)$
\[
\begin{array}{cccc}
h: & \mathbb{R}^{N} & \rightarrow & \mathbb{R}^{P}\\
& \widehat{\theta} & \mapsto & \widehat{\ell}
\end{array}
\]

Our aim is to search these precoding and decoding funtion that can minimize the average utility loss
\begin{equation}
\mathbb{E}_{\ell}\left[\left|u\left(x^{\star}\left(\ell\right);\ell\right)-u\left(x^{\star}\left(\widehat{\ell}\right);\ell\right)\right|^2\right].\label{eq:top}
\end{equation}

Therefore, the goal is to choose the best data pre-processing schemes $g$ and $h$ to minimize the above optimality loss. Note that the conventional paradigm consisting in minimizing distorsion is readily obtained as a special instance by choosing $u$ to be the identity function.

\section{Linear approximation of the optimal decision}
\label{sec:Solution-for-a-fixed-l}

\subsection{Solution of OP for fixed data}
\label{ssec:solution}
For a given realization of random vector $\boldsymbol{\ell}\in\mathbb{R}^{P}$ , we
want to find $x^{\star}(\boldsymbol{\ell})$ defined in \eqref{eq:xstar}. Without loss of generality, we assume
\begin{equation}
\ell_{1}\leqslant\ell_{2}\leqslant\dots\leqslant\ell_{z}<0\leqslant\ell_{z+1}\leqslant\dots\leqslant\ell_{P}\label{eq:assumption}
\end{equation}

When $p\in\mathbb{N}^{+}$, $x^{\star}\left(\ell\right)  \in\arg\min_{x}\left(\sum_{j=1}^{P}|x_{j}+\ell_{j}|^{p}\right)^{\frac{1}{p}}$ which is equivalent to $x^{\star}\left(\ell\right)  \in\arg\min_{x}\sum_{j=1}^{P}|x_{j}+\ell_{j}|^{p}$. And we can discuss in two cases: $0<E<\sum_{j=1}^{z}|\ell_{j}|$ or $E\geqslant\sum_{j=1}^{z}|\ell_{j}|$.

In the first case, a solution for this OP should satisfy
\begin{equation}
\sum_{j=1}^{z}x^{\star}_{j}=E,\ x^{\star}_{z+1}=\dots=x^\star_{P}=0
\label{eq:cond1}
\end{equation}
and 
\begin{equation}
x^\star_j+l_j<0,\ \forall j\in\{1,2,\dots,z\}
\label{eq:cond2}
\end{equation} 

Because, for \eqref{eq:cond1}, if $\exists j\in\{z+1,\dots,P\},\ x^\star_{j}>0$, for sure, $\exists i\in\{1,\dots,z\}$, $x^\star_{i}+\ell_{i}<0$. We can use $x^\dagger_j=x^\star_{j}-\min\left(x^\star_{j},-x^\star_{i}-\ell_i\right)$ and $x^\dagger_i=x^\star_{i}+\min\left(x^\star_{j},-x^\star_{i}-\ell_i\right)$ replacing respectively $x^\star_{j}$ and $x^\star_{i}$, obtaining a p-norm less than the original which implies $x^\star_{j}$ and $x^\star_{i}$ are not the solutions.

The reason for having \eqref{eq:cond2} is similar. If $\exists j\in\{1,2,\dots,z\},\ x^\star_{j}+\ell_{j}>0$, and for sure, $\exists i\in\{1,\dots,z\}$, $x^\star_{i}+\ell_{i}<0$ We can obtain a better result by replacing $x^\star_{j}$ and $x^\star_{i}$ with $x^\dagger_j=x^\star_{j}-\min\left(x^\star_{j}+\ell_{j},-x^\star_{i}-\ell_i\right)$ and $x^\dagger_i=x^\star_{i}+\min\left(x^\star_{j}+\ell_{j},-x^\star_{i}-\ell_i\right)$ respectively.

And from the OP which is easy to prove to be convex, we can thus use KKT and get Lagrangian function $\mathcal{L}$
\begin{equation}
\mathcal{L}=\sum_{j=1}^{z}\left(-x_{j}-\ell_{j}\right)^{p}+\delta\left(E-\sum_{j=1}^{z}x_{j}\right)-\sum_{j=1}^{z}\lambda_{j}x_{j}
\end{equation}
\begin{equation}
\frac{\partial\mathcal{L}}{\partial x_{j}}=-p\left(-x_{j}-\ell_{j}\right)^{p-1}-\delta-\lambda_{j}
\end{equation}

\begin{equation}
\begin{cases}
-p\left(-x_{j}^{\star}-\ell_{j}\right)^{p-1}-\delta^{\star}-\lambda_{j}^{\star} & =0\\
E-\sum_{j=1}^{P}x_{j}^{\star} & =0\\
-x_{j}^{\star} & \leqslant0\\
\lambda_{j}^{\star}x_{j}^{\star} & =0
\end{cases}
\end{equation}

Getting rid of the slack variables $\left\{ \lambda_{j}^{\star}\right\} $,
we have

\begin{equation}
-p\left(-x_{j}^{\star}-\ell_{j}\right)^{p-1}=\delta^{\star}
\end{equation}
\begin{equation}
x_{j}^{\star}=\left(-\left(-\frac{\delta^{\star}}{p}\right)^{1-p}-\ell_{j}\right)^{+}
\end{equation}
where $\left(a\right)^{+}=\max(a,0)$. This is a water-filling solution.
We assume $\mu=-\left(-\frac{\delta^{\star}}{p}\right)^{1-p}$ is water
level, thus

\begin{equation}
x_{j}^{\star}=\left(\mu-\ell_{j}\right)^{+}\label{eq:waterload}
\end{equation}
according to \eqref{eq:assumption} $\mu$ is obtained by

\begin{equation}
\mu=\frac{1}{n^{\star}}\left(E+\sum_{j=1}^{n^{\star}}\ell_{j}\right)\label{eq:waterlevel}
\end{equation}
where $n^{\star}$ indicates the number of induces that should be loaded
\begin{equation}
n^{\star}=\max_{n\leqslant z}\left\{ n:\left(n-1\right)\ell_{n}-\sum_{j=1}^{n-1}\ell_{j}<E\right\} 
\end{equation}

In the second case, a solution should satisfy
\begin{equation}
x^\star_{j}+\ell_{j}>0,\ \forall j\in\{1,\dots,P\}
\end{equation}
the reason is similar to \eqref{eq:cond2}. Thus, we get a new Lagrangian function $\mathcal{L}$
\begin{equation}
\mathcal{L}=\sum_{j=1}^{P}\left(x_{j}+\ell_{j}\right)^{p}+\delta\left(E-\sum_{j=1}^{P}x_{j}\right)-\sum_{j=1}^{P}\lambda_{j}x_{j}
\end{equation}

But we can get also a water-filling solution by the similar way to the first case.

When $p\rightarrow+\infty$, this OP can also be solved by a water-filling solution obtained by a similar way when $p\in \mathbb{N}^{+}$. Due to the limit of space, the discussion is omitted here.

\subsection{Linear approximation of the water-filling operator around fixed data}
\label{ssec:linear_approximation}

Water-filling solution a widely used in communication systems \cite{4036033}-\cite{5447064}. One has the solution $x^{\star}\left(\ell\right)$, but it is hard to have a explicit expression of this solution. In order to facilitate the operation on water-filling solution, like computing the gradient, our aim in this section is to find a linear approximation of this solution. Its Taylor series expansion can be expressed as
\begin{equation}
\begin{aligned}
x^{\star}\left(\ell+d\ell\right)&=x^{\star}\left(\ell\right)+\left(\nabla_{\ell}x^{\star}\left(\ell\right)\right)^{T}d\ell+o\left(d\ell\right)\\
&=x^{\star}\left(\ell\right)+\boldsymbol{H}d\ell+o\left(d\ell\right)
\end{aligned}
\end{equation}
where $\boldsymbol{H}$ is the Jacobian matrix.

From the discussion of the minimization problem of $||x+\ell||_{p}$
with different $p$. $x^{\star}\left(\ell\right)=\left[x_{1}^{\star},x_{2}^{\star},\dots,x_{P}^{\star}\right]^{T}$
is the function of $\ell$. We consider water-filling
solution in a linear region where the nonzero/zero elements is still nonzero/zero. $n^{\star}$ thus is fixed. Accoding to the admissible values of $n^{\star}$ ($n^{\star}=1,\dots,P$), $\mathbb{R}^{P}$ can be divided into several regions
$$
\mathcal{I}\subseteq\{1,2,\dots,P\},\ \mathcal{I}\neq\varnothing
$$
$$
\mathcal{M}_{\mathcal{I}}=\{\ell|x^{\star}_k\left(\ell\right)>0,\ k\in\mathcal{I},\ x^{\star}_j\left(\ell\right)=0,\ j\notin\mathcal{I}\}
$$
$\mathcal{I}$ is powerset of $\{1,2,\dots,P\}$ excluding the empty set. $\mathbb{R}^{P}$ is divided into $2^P-1$ regions
$$
\forall i\neq j,\ \mathcal{M}_i\bigcap\mathcal{M}_j=\varnothing,\ \bigcup_{i=1}^{2^P-1}\mathcal{M}_i=\mathbb{R}^{P}
$$

In each region, we can have a linear approximation of water-filling operator. It means that $\boldsymbol{H}$ is different in different regions. $\boldsymbol{H}$ is thus a function of $\ell$, denoted by $\boldsymbol{H}\left(\ell\right)$.

Consider one type of these regions ($\boldsymbol{H}\left(\ell\right)$ can be obtained by the same way in other regions) where we assume last $\left(P-n^{\star}\right)$ elements of $\ell$
will not affect on the output, $x_{n^{\star}+1}^{\star}=x_{n^{\star}+2}^{\star}=\cdots=x_{P}^{\star}=0$.
We focus on the influence of the first $n^{\star}$ elements of $\ell$.
We consider all admissible values of the first $n^{\star}$ elements
that satisfy $n^{\star}$ fixed and
\begin{equation}
\ell_{1},\dots,\ell_{n^{\star}}<\ell_{n^{\star}+1}\leqslant\dots\leqslant\ell_{P}
\end{equation}

Thus from Equations \eqref{eq:waterload} and \eqref{eq:waterlevel}, we have
\begin{equation}
x_{j}^\star=
\begin{cases}
\frac{\sum_{i=1}^{n^{\star}}\ell_{i}+E}{n^{\star}}-\ell_{j}& j\leqslant n^{\star}\\
0 & j>n^{\star}\\
\end{cases}
\end{equation}

By calculating the gradient of $x^\star\left(\ell\right)$, we have
\begin{equation}
\boldsymbol{H}\left(\ell\right)=\left(\begin{array}{ccccccc}
-1+\frac{1}{n^{\star}} & \frac{1}{n^{\star}} & \text{\ensuremath{\cdots}} & \frac{1}{n^{\star}} & 0 & \cdots & 0\\
\frac{1}{n^{\star}} & -1+\frac{1}{n^{\star}} & \cdots & \vdots & \vdots &  & \vdots\\
\vdots & \vdots &  & \vdots & \vdots &  & \vdots\\
\frac{1}{n^{\star}} & \cdots & \cdots & -1+\frac{1}{n^{\star}} & 0 & \cdots & 0\\
0 & \cdots & \cdots & 0 & 0 & \cdots & 0\\
\vdots &  &  & \vdots & \vdots &  & \vdots\\
0 & \cdots & \cdots & 0 & 0 & \cdots & 0
\end{array}\right)
\end{equation}
and a constant term is obtained easily
\begin{equation}
b\left(\ell\right)=\left(\begin{array}{c}
\frac{E}{n^{\star}}\\
\frac{E}{n^{\star}}\\
\vdots\\
\frac{E}{n^{\star}}\\
\vdots\\
0
\end{array}\right)
\end{equation}

For the same reason as $\boldsymbol{H}\left(\ell\right)$, $b$ depends on $\ell$ and is denoted
by $b\left(\ell\right)$.

\section{Best Linear Transformation}
\label{sec:linear_transformation}

Linear transformation is a classic approximation that project the signal to the chosen basis. And the signal is represented by a few of the vectors in this basis. Karhunen-Loève Transformation (KLT) have been proved be an optimal approximation in the sense of mean-square error \cite{STEPHANE2009435}. The encoding and decoding scheme can be considered as functions $g\left(\cdot\right)$ and $h\left(\cdot\right)$. For simplification,
we consider an linear approximation without quantizer, and $\theta=g\left(\ell\right)=\boldsymbol{B}\ell$,
$\widehat{\ell}=h\left(\theta\right)=\boldsymbol{B}^{T}\boldsymbol{B}\ell$, where $\boldsymbol{B}\in\mathbb{R}^{N\times P}$. \eqref{eq:top} becomes

\begin{equation}
E_{\ell}\left[\left|u\left(x^{\star}\left(\ell\right),\ell\right)-u\left(x^{\star}\left(\boldsymbol{B}^{T}\boldsymbol{B}\ell\right),\ell\right)\right|^2\right]\label{eq:opt_loss}
\end{equation}

In the case \eqref{eq:opt_loss}, we search a linear
mapping from $\mathbb{R}_{+}^{P}$ to $\mathbb{R}^{P}$ with maximum rank
$N$.

With a given set of realizations of $\ell$, $L=\{\ell^{\left(1\right)},\ell^{\left(2\right)},\dots,\ell^{\left(T\right)}\}$, the loss function is equivalent to
\begin{equation}
\begin{aligned}\Gamma\left(\boldsymbol{B}\right) =\frac{1}{T}\sum_{i=1}^{T}&\left(u\left(x^{\star}\left(\ell^{\left(i\right)}\right),\ell^{\left(i\right)}\right)-u\left(x^{\star}\left(\widehat{\ell}^{\left(i\right)}\right),\ell^{\left(i\right)}\right)\right)^{2}\\
=\frac{1}{T}\sum_{i=1}^{T}&\left(u\left(x^{\star}\left(\ell^{\left(i\right)}\right),\ell^{\left(i\right)}\right)+||x^{\star}\left(\widehat{\ell}^{\left(i\right)}\right)+\ell^{\left(i\right)}||_{p}\right)^{2}\\
=\frac{1}{T}\sum_{i=1}^{T}&\left({u\left(x^{\star}\left(\ell^{\left(i\right)}\right),\ell^{\left(i\right)}\right)}\right.\\
&\left.{+||\left(\boldsymbol{H}_{i}\boldsymbol{B}^{T}\boldsymbol{B}+\boldsymbol{I}\right)\ell^{\left(i\right)}+b_{i}||_{p}}\right)^{2}
\end{aligned}
\label{eq:yb}
\end{equation}
Denote $\boldsymbol{H}\left(\widehat{\ell}^{\left(i\right)}\right)$
and $b\left(\widehat{\ell}^{\left(i\right)}\right)$, the linear approximation parameters of water-filling operator, 
by $\boldsymbol{H}_{i}$, $b_{i}$ respectively, and assume that they are
independent of $\boldsymbol{B}$ for simplification.

We consider gradient descent method searching an optimal $\boldsymbol{B}^{*}$.
\begin{algorithm}
	\caption{Gradient descent searching $\boldsymbol{B}^{*}$} 
	\label{alg1}
	\begin{algorithmic}
		\REQUIRE Initial matrix $\boldsymbol{B}$ (KL basis)
		\renewcommand{\algorithmicrequire}{\textbf{Input:}}
		\REQUIRE Learning rate $\epsilon=0.05$
		\WHILE{$iteration_{max}$ not reached and optimality loss reduced more than $0.01\%$} 
		\STATE Compute gradient: $\boldsymbol{G}\leftarrow\nabla_{\boldsymbol{B}}\Gamma\left(\boldsymbol{B}\right)$
		\STATE Apply update: $\boldsymbol{B}=\boldsymbol{B}-\epsilon \boldsymbol{G}$ 
		\ENDWHILE
	\end{algorithmic}
\end{algorithm}

Even if our problem is not convex, we can find a local optimal which is verified by experiment.

When $p\in\mathbb{N}^{+}$, after some calculation, we can get the gradient

\begin{equation}
\nabla_{\boldsymbol{B}}\Gamma\left(\boldsymbol{B}\right)=\frac{1}{T}\sum_{i=1}^{T}C_{i}\left(\boldsymbol{B}\ell^{\left(i\right)}\beta_{i}^{T}\boldsymbol{H}_{i}+\boldsymbol{B}\boldsymbol{H}_{i}^{T}\beta_{i}\ell^{\left(i\right)T}\right)\label{eq:ybd}
\end{equation}
where
\begin{equation}
\begin{aligned}
C_{i}=2&\left({u\left(x^{\star}\left(\ell^{\left(i\right)}\right),\ell^{\left(i\right)}\right)}\right.\\
&\left.{-u\left(x^{\star}\left(\widehat{\ell}^{\left(i\right)}\right),\ell^{\left(i\right)}\right)}\right)||x^{\star}\left(\widehat{\ell}^{\left(i\right)}\right)+\ell^{\left(i\right)}||_{p}^{1-p}
\end{aligned}
\end{equation}
\begin{equation}
\beta_{i}=\text{\ensuremath{\underbrace{\left(x^{\star}\left(\widehat{\ell}^{\left(i\right)}\right)+\ell^{\left(i\right)}\right)\odot\cdots\odot\left(x^{\star}\left(\widehat{\ell}^{\left(i\right)}\right)+\ell^{\left(i\right)}\right)}_{p-1}}}
\end{equation}

When $p\rightarrow+\infty$, the gradient is obtained by
\begin{equation}
\nabla_{\boldsymbol{B}}\Gamma\left(\boldsymbol{B}\right)=\frac{1}{T}\sum_{i=1}^{T}D_{i}\left(\boldsymbol{B}\ell^{\left(i\right)}s_{k\left(i\right)}^{T}\boldsymbol{H}_{i}+\boldsymbol{B}\boldsymbol{H}_{i}^{T}s_{k\left(i\right)}\ell^{\left(i\right)T}\right)\label{eq:ybdi}
\end{equation}
where
\begin{equation}
D_{i}=2\left(u\left(x^{\star}\left(\ell^{\left(i\right)}\right),\ell^{\left(i\right)}\right)-u\left(x^{\star}\left(\widehat{\ell}^{\left(i\right)}\right),\ell^{\left(i\right)}\right)\right)
\end{equation}
\begin{equation}
s_{k\left(i\right)}=\left(\begin{array}{c}
0_{(k\left(i\right)-1)\times1}\\
1\\
0_{(P-k\left(i\right))\times1}
\end{array}\right)
\end{equation}
assume for $x^{\star}\left(\widehat{\ell}^{\left(i\right)}\right)+\ell^{\left(i\right)}$,
$k\left(i\right)$-th element has the maximum value.





\section{Best Nonlinear Transformation}
\label{sec:nonlinear_transformation}

The errors of approximation depends on the signal. Linear and nonlinear approximation have a similar performance if signal is uniformly regular. However, for the nonuniformly regular signal, nonlinear approximation performs better \cite{STEPHANE2009435}. We thus consider also that function $g\left(\cdot\right)$ is nonlinear.
Because of the quantizer, we still consider $h\left(\cdot\right)$
as a linear function.
\begin{figure}[h]
	\includegraphics[scale=0.38]{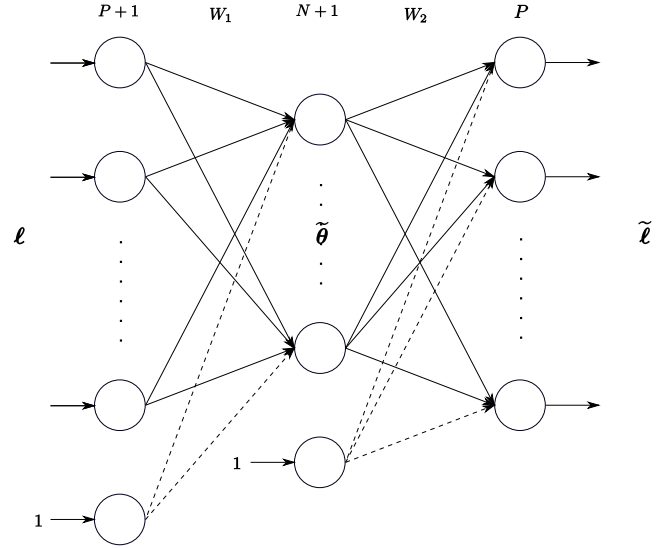}
	
	\caption{Structure of NN}
	\label{fig:AUTO}
\end{figure}

In this case, we propose to solve \eqref{eq:top} by using autoencoder \cite{Goodfellow-et-al-2016}. An autoencoder is a type of unsupervised learning neural network. This network is trained to copy its input $\ell$ to its output $\widetilde{\ell}$ in order to efficiently encode the data. Autoencoder is typically used for dimensionality reduction or feature learning. The network can be considered as unite of two parts. One part, for encoding, is trained to ignore the piddling information, the dimension of the input thus can be reduced. Another part, for decoding, is trained to reconstruct, from the reduced encoded parameters $\widetilde{\theta}$, an output that minimizes the loss function (usually considered as mean-square error).

Fig. \ref{fig:AUTO} illustrate the structure of autoencoder that we used. With reduction side, the network encodes from input $\ell$ to $\widetilde{\theta}$ whose dimension $N$ is much less than the dimension of the original input. The reconstructing side generate an output $\widetilde{\ell}$ from $\widetilde{\theta}$. 

Denoting $W_{i,j}^{(\gamma)}$ the weight between the neuron $j$ in the $\gamma$-th layer and the neuron $i$ in $(\gamma+1)$-th layer. And $o_j^{(\gamma)}$ is the output of $j$-th neuron in $\gamma$-th layer, $b_i^{(\gamma)}$ is the bias term for neuron $i$, and this term is indicated by $1$ in the Fig. \ref{fig:AUTO}. The basic model for autoencoder is given by:
$$
o_i^{(\gamma+1)}=f\left(b_i^{(\gamma)}+\sum_{j=1}^{N_{\gamma}}W_{i,j}^{(\gamma)}o_j^{(\gamma)}\right)
$$
where $N_{\gamma}$ indicate the number of neuron in $\gamma$-th layer, $f\left(\cdot\right)$ is the activation function. After some matrix manipulations, we get the expressions of $\widetilde{\theta}$ and $\widetilde{\ell}$
\begin{equation}
\widetilde{\theta}=f\left(\boldsymbol{W}_{1}\zeta\right)
\end{equation}
\begin{equation}
\widetilde{\ell}=\boldsymbol{W}_{2}\eta
\end{equation}
where
$$
\zeta=\left(\begin{array}{c}
\ell\\
1\\
\end{array}\right)
$$
$$
\eta=\left(\begin{array}{c}
\widetilde{\theta}\\
1\\
\end{array}\right)
$$
their entries $1$ are for bias term, and
$$
\boldsymbol{W}_{1}=\left(\begin{array}{cccc}
W_{1,1}^{\left(1\right)} & \cdots & W_{1,P}^{\left(1\right)} & b_1^{\left(1\right)} \\
\vdots& &\vdots&\vdots\\
W_{N,1}^{\left(1\right)} & \cdots & W_{N,P}^{\left(1\right)} & b_N^{\left(1\right)} \\
\end{array}\right)
$$
$$
\boldsymbol{W}_{2}=\left(\begin{array}{cccc}
W_{1,1}^{\left(2\right)} & \cdots & W_{1,N}^{\left(2\right)} & b_1^{\left(2\right)} \\
\vdots& &\vdots&\vdots\\
W_{P,1}^{\left(2\right)} & \cdots & W_{P,N}^{\left(2\right)} & b_P^{\left(2\right)} \\
\end{array}\right)
$$

Instead of mean-square error, in our autoencoder, we define a loss function depending on the utility loss
\begin{equation}
\left(||x^{\star}\left(\widetilde{\boldsymbol{\ell}}\right)+\boldsymbol{\ell}||_p-||x^{\star}\left(\boldsymbol{\ell}\right)+\boldsymbol{\ell}||_p\right)^2
\end{equation}

\section{Numerical performance analysis}
\label{sec:Simulation}
In this section, we present some simulation results. These results illustrate the performance of KLT, proposed approaches: linear transformation and nonlinear transformation.

Our dataset comes from Ausgrid \cite{ausgrid}. Consider one user's daily energy consumption as a random vector $\ell$. We choose the energy consumption of one user during one year as data set. The data are measured each half hour, and the energy consumption is transmitted once each day. Thus $P=48$ and the data set is of dimension $48\times 356$. The rank limit is $N=4$. Other parameters is set as $p=\inf$ or $20$, $E=50$. 


For linear transformation, we take KL basis as initial value of $\boldsymbol{B}$. Alg. \ref{alg1} is then executed.

We choose one hidden layer neural network for nonlinear transformation. Since the rank limit is 4, hidden layer has 4 neurons. The outputs of these 4 neurons are exactly the encoded information. Since sigmoid function restrict its output between $0$ and $1$, in order to simplify the quantization, we choose it as activation function.
\begin{figure}[htbp]
	
	\textsf{\includegraphics[scale=0.23]{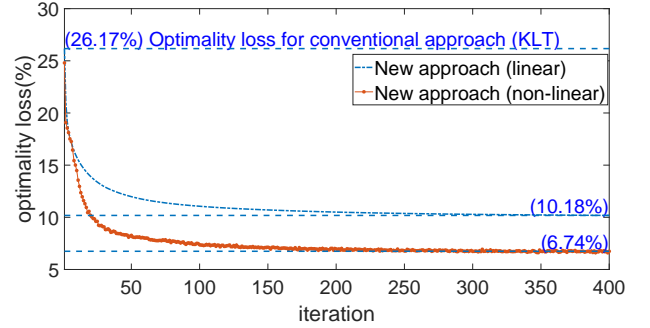}}
	\par
	\caption{When $p \rightarrow\infty$, it is seen that the optimality loss induced by the compression noise can be reduced in typical scenarios from about $25\%$ (KLT performance) to about $6\%$ (new approach).}
	\label{fig:E50e}
\end{figure}

\begin{figure}[htbp]
	
	\textsf{\includegraphics[scale=0.23]{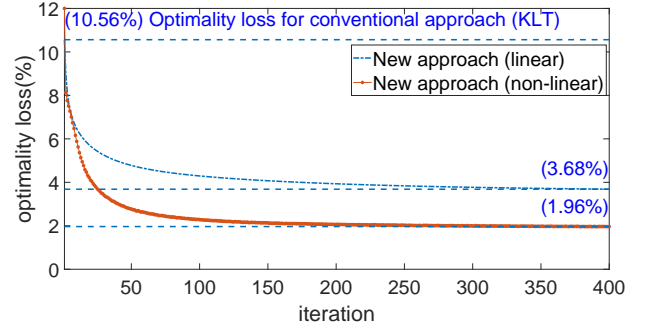}}
	\par
	\caption{When $p=20$, the impact of compression noise is generally less than for $p \rightarrow\infty$. A careful design allows one to obtain very small optimality losses.}
	\label{fig:E50ep20}
\end{figure}

Fig. \ref{fig:E50e}-\ref{fig:E50ep20} illustrate that, with different value of $p$, both linear and nonlinear transformation reduce efficiently the optimality loss and have a better performance than KLT. Compared to linear transformation, the nonlinear one is better. As we introduced in Section \ref{sec:nonlinear_transformation}, nonlinear approximation is more suitable for nonuniformly regular data. By our nonlinear transformation, the data that we used or other data which have some similar characteristics can take advantage of nonlinear approximation.

\begin{figure}[htbp]
	\begin{centering}
		\textsf{\includegraphics[scale=0.23]{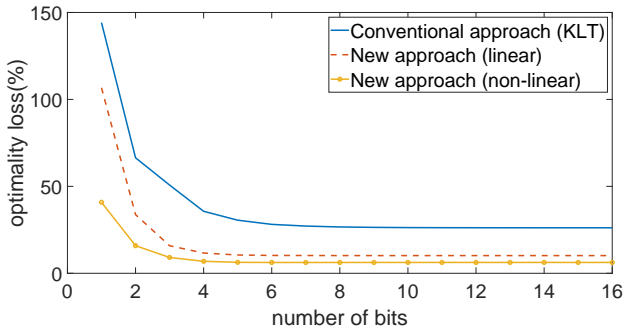}}
		\par\end{centering}
	\caption{For $p \rightarrow\infty$, this curve exhibits the rate-optimality loss tradeoff curve. For rate above 6 bits/sample, improving the compression quality has no impact on the final utility function maximization.}
	\label{fig:E50uq}
\end{figure}

\begin{figure}[htbp]
	\begin{centering}
		\textsf{\includegraphics[scale=0.23]{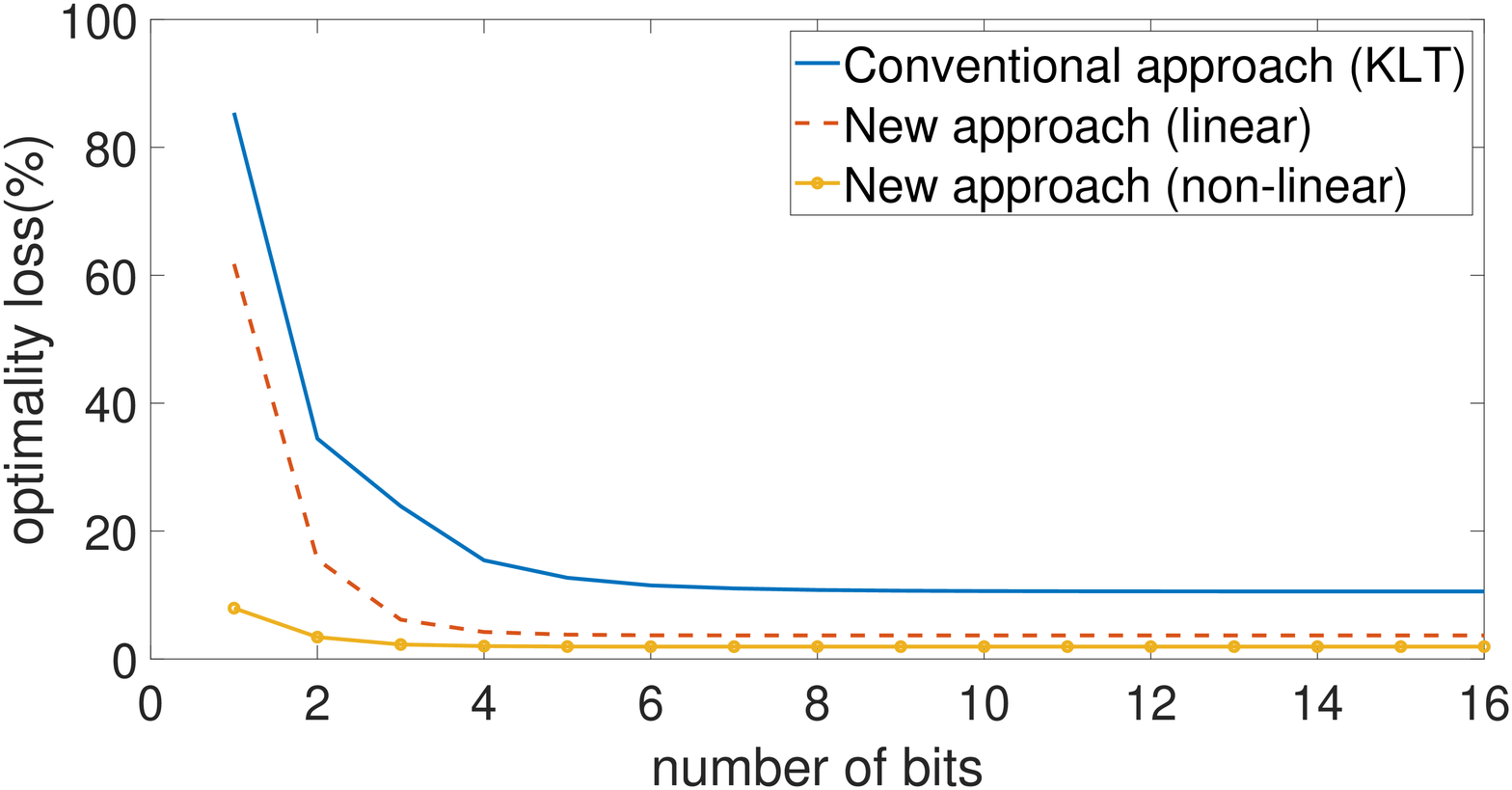}}
		\par\end{centering}
	\caption{For $p=20$ the previous observation holds.}
	\label{fig:E50uqp20}
\end{figure}

A dead-zone uniform quantizer follows after the precoding process. The quantized coefficients are transmitted losslessly. We thus obtain Fig. \ref{fig:E50uq}-\ref{fig:E50uqp20} which illustrate the optimality loss as function of bits limitation of the quantizer. Even with only one bit, the linear and nonlinear transformations have a great improvement compared to conventional approach (KLT) in the sense of optimality loss. In the low bit rate region, the optimality loss reduced significantly with the increasing bit limitation. And the curves tend to flat in high bit rate region. The results shows that, by our method, even only 4 bits allocated to each encoded coefficients (16 bits for a realization of $\ell$) are sufficient to ensure an appropriate optimality loss. Even if we allocate more bits, the improvement of optimality loss is not distinct.

\section{Conclusion}
\label{sec:conclusion}

In this paper, a new perspective of data pre-processing (or precoding) is taken. We present a use-oriented formulation of the problem and derive two use-oriented pre-processing schemes (namely, a linear and a non-linear scheme). These two new schemes are compared with conventional data pre-processing schemes such as the KLT. By using real smart grid data, numerical experiments show that, in the sense of the optimality loss, linear and nonlinear transformations perform much better than conventional schemes such as the KLT. For instance, the optimality loss induced by the compression noise can be reduced in typical scenarios from about $25\%$ to about $6\%$.

\end{document}